\documentclass[figures]{epl}
\usepackage{epsfig}
\title{Scaling and hydrodynamic effects in lamellar ordering}
\shorttitle{Lamellar ordering}
\author{Aiguo Xu\inst{1}\thanks{Present address: Dept. of Physics,
Yoshida-south Campus, Kyoto University, Sakyo-ku, Kyoto, 606-8501 Japan} 
\and G. Gonnella\inst{1} \and A. Lamura\inst{2}
\and G. Amati\inst{3} \and F. Massaioli\inst{3}}
\shortauthor{Aiguo Xu \etal}
\institute{
     \inst{1} Dipartimento di Fisica, Universit\`a di Bari {\rm and}
 Center of Innovative Technologies for Signal Detection and Processing
 (TIRES) {\rm and} INFN (Istituto Nazionale di Fisica Nucleare), Sez. di  Bari
{\rm and} Istituto Nazionale per la  Fisica della Materia,\\ 
via Amendola 173, 70126 Bari, Italy\\
     \inst{2} Istituto Applicazioni Calcolo, CNR, Sezione di Bari,\\ 
              via Amendola 122/D, 70126 Bari, Italy\\
     \inst{3} CASPUR, Via dei Tizii 6/b, 00185 Rome, Italy
}
\pacs{64.75.+g}{Solubility, segregation, and mixing; phase separation}
\pacs{05.70.Ln}{Nonequilibrium and irreversible thermodynamics}
\pacs{47.20.Hw}{Morphological instability; phase changes}
\pacs{82.35.Jk}{Copolymers, phase transitions, structure}

\begin{document}

\maketitle

\begin{abstract}
We study the kinetics of domain growth of  fluid mixtures quenched
from a disordered to a lamellar phase. At low viscosities, in two
dimensions, when
 hydrodynamic modes become important, dynamical scaling is verified in the form
 $C(\vect{k}, t) \sim L^{\alpha} f[(k-k_M)L]$ where $C$ is the
 structure factor with maximum at $k_M$ and
 $L$ is a typical length of the system with
 logarithmic growth at late times.
 The presence of extended defects can explain the behavior of $L$.
Three-dimensional simulations confirm that diffuse grain boundaries
inhibit complete ordering of lamellae.
Applied shear flow  alleviates frustration and
 gives  power-law growth  at all times.
\end{abstract}

\section{Introduction}
The kinetics of the growth of ordered phases after a quench from a
disordered state continues to provide interesting physical
questions. While the process is reasonably well understood for the
case of binary mixtures, where dynamical scaling occurs and
domains grow with power-law behavior \cite{B94}, basic questions
remain to be clarified  for more complex systems \cite{Y01}. In
this Letter we consider the case of fluid mixtures where, due to
competing interactions, the system would arrange itself in
stripes. Examples are di-block copolymer melts, where chains of
type A and B, covalently bonded end-to-end in pairs, segregate at
low temperatures in regions separated by a stack of lamellae
\cite{Bat}, or   ternary mixtures where surfactant form interfaces
between oil and water \cite{GS94}. Other systems with lamellar
order are smectic liquid crystals \cite{Deg}, dipolar \cite{SD}
and supercooled liquids \cite{KKZNT95}, chemically reactive binary
mixtures \cite{GC}.
 Lamellar patterns are also observed in Raleigh-B\'enard cells 
above the convective
threshold \cite{CH}.

The ordering of lamellar systems is characterized by the presence
of frustration on large scales. This affects the late time
evolution  which, as discussed below, can be very slow or also
frozen. Relations with the dynamics of structural glasses have
been also considered in recent literature \cite{BV02,SW00}. One
can expect that in real systems the effects of the velocity field,
inducing motion around local or extended defects, can be relevant.
Our purpose here is to analyze the process of lamellar ordering
taking into account the effects of hydrodynamics.
We consider a model
based on a Ginzburg-Landau free-energy, where dynamics is
described by  convection-diffusion and Navier-Stokes equations \cite{yeo}.
Our results for two-dimensional systems show that, when
hydrodynamical modes are effective (at sufficiently low
viscosities), dynamical scaling holds  with a complex dependence
of characteristic lengths on time, analogous  to that of systems
with quenched disorder \cite{RC} not present here. An intermediate
faster growth is followed by a slower logarithmic evolution at
later times.
The  slowing-down of the dynamics is attributed to the formation
of extended
defects between domains of  lamellae
with different orientation, almost  perpendicular each other.
Our simulations  show that the same phenomenon occurs in
three-dimensional systems.
These results are not expected to depend on the specific model
considered and could be relevant for a broad class of systems with
lamellar order. We  also verified  that  applied shear flows
alleviate frustration giving power-law growth   at all times.

\section{The model}
We consider the free-energy model \cite{B75,Lei80}
\begin{equation}
{\cal F}\{\varphi\} = \int d^d x
\{\frac{a}{2} \varphi^2 + \frac{b}{4} \varphi^4
+ \frac{\kappa}{2} \mid \nabla \varphi \mid^2 +\frac {c}{2} (\nabla^2
\varphi)^2\}
\label{eqn1}
\end{equation}
where  $\varphi$ is the order parameter field representing the
local concentration difference between the two components of the
mixture. The parameters $b$ and $ c $ have to be positive in order
to ensure stability.  For  $a<0$ and $\kappa
> 0 $  the  two homogeneous phases with $\varphi = \pm \sqrt{-a/b}$ coexist.
A negative $\kappa$ favors the presence of interfaces and a
transition into a lamellar phase can occur.  In single mode
approximation,
 assuming a profile like $A \sin k_0 x$ for the direction transversal to the 
lamellae,
 one finds   the transition ($|a|=b$) at  $ a \approx - 1.11 \kappa^2 /c $
 where  $ k_0 = \sqrt{ -\kappa /2c }$, $A^2 = 4 (1 +
\kappa^2/4cb)/3$.
 The expression (1) can be
also written for negative $\kappa$ as ${\cal F}\{\varphi\} = \int
d^d x \{\frac{\tau}{2} \varphi^2 + \frac{b}{4} \varphi^4 +
\frac{c}{2}[(\nabla ^2 + k_0^2) \varphi)^2]\}$,  $\tau = a - c
k_0^4 $, and in this form is generally  used to describe di-block
copolymers in the weak segregation limit\cite{Lei80}.

The dynamical equations are \cite{Y01}
\begin{equation}
\frac {\partial  v_{\alpha}}{\partial t} + \vect{v} \cdot  \vect{\nabla}
 v_{\alpha} = - \frac{1}{\varrho} \frac {\partial P_{\alpha
\beta}} {\partial x_{\beta}} + \nu \nabla^2 v_{\alpha}  \quad,
\label{motion1}
\end{equation}
\begin{equation}
\frac {\partial \varphi}{\partial t}+ \vect{\nabla} \cdot (\varphi
\vect{v}) = \Gamma \nabla^2 \frac {\delta {\cal F}} {\delta \varphi}
\quad, \label{motion2}
\end{equation}
where  $v_{\alpha} $
 are the components
of the velocity field, $\varrho$ is the total density  of the
mixture and the incompressibility condition $\vect{\nabla} \cdot
\vect{v} = 0 $ is considered. $\nu $ is the kinematic viscosity and
$\Gamma$ is a mobility coefficient. The pressure tensor is the sum
of the usual hydrodynamical part and of a tensor $P_{\alpha
\beta}^{chem} $ depending on $ \varphi$, with a functional form
obtainable from the free energy  containing off-diagonal terms,
whose expression can be found in \cite{yeo}.  It can be shown that
$\delta \mathcal {F} /\delta \varphi = 0 $ implies
$\partial_{\alpha} P_{\alpha \beta}^{chem} = 0$. The laplacian in
the r.h.s. of Eq.~(\ref{motion2}) ensures the conservation of the
order parameter. We will simulate these equations by using   a finite
difference scheme for the convection-diffusion equation and a
Lattice Boltzmann Method (LBM) for  the Navier-Stokes equation.
Advantages of this method \cite{det} with respect to other  LBM
schemes for fluid mixtures are that spurious terms
 appearing in Eq.(3) \cite{Swift96} are avoided
and  the numerical efficiency is increased. In $D=3$ we  used a
parallel version of the LBM code which {\it fuses} the
streaming and the collision processes to reduce bandwidth
requirements \cite{giorgio}.

\begin{figure}[t]
\begin{center}
Fig.~1\hspace{0.5\textwidth}Fig.~2
\end{center}
\vskip -0.5cm
\caption{Configurations at
different times of a $256 \times 256$ portion of the lattice.}
\caption{Time behavior of characteristic size $L$
and $C_M \equiv C(k_M,t)$ on log-log (inset) and log-linear
scales. The straight line in the inset has slope $0.3$.}
\end{figure}

We can summarize now what is known on the ordering properties of
lamellar fluids. Previous two-dimensional studies of the full
model (1-3) \cite{yeo}, without quantitative analysis due to the
small size of the lattice considered ($128^2$), showed  the
relevance of hydrodynamics  for obtaining well ordered domains on
the scale of the system simulated. More results exist for variants
of the model (1) with long-range interactions, without
hydrodynamics \cite{LG89,STY96,MZF98} or neglecting the inertial
terms of the l.h.s. of Eq.(2) \cite{YS02}. Eq.(3) without
advection and with non-conserved order parameter,
 corresponding to the  Swift-Hohenberg model for
Raleigh-B\'enard convection \cite{SH,EVG92,HSG97,CM95,BV01}, has
been also largely studied. Regimes with dynamical scaling have
been found with the order parameter structure factor behaving as
 $ C(\vect{k}, t) \equiv <|\varphi_{\vect{k}}(t)|^2> 
\sim t^z f[(k - k_0) t^z]$ with different values for the exponent $z$,
 $ \varphi_{\vect{k}}(t) $ being the Fourier transform of 
$\varphi(\vect{x}, t)$
\cite{EVG92,CM95,HSG97,CB98,QM02,YS02,Har00,BV01}.
Slower evolution with frozen states and grain boundary pinning has
been also observed \cite{BV02,HSG97} for very deep quenches. In $D=3$
the few existing simulations have not considered dynamical scaling 
\cite{coveney,ohta}.

\section{Ordering dynamics}
We run our two-dimensional simulations on
a $1024^2$  lattice starting from disordered configurations. We
fixed $|a| = b $ so that the minimum of the potential part of the
free-energy is at $\varphi = \pm 1$. We checked on small systems
($128^2$) for different $\kappa $ in the lamellar phase that the
expected equilibrium state was reached. The
value of $c$ was fixed in such a way that the period is about 10
times the lattice spacing and $|a| = 2 \times 10^{-4}, \kappa = -3
\times |a|, c = 3.8 \times |a| , \Gamma = 25$
for the cases presented. Details on the LBM part
of the code can be found in \cite{Xu03}.

After the quench,  lamellae start to form  evolving until the
equilibrium wavelength is locally reached. This part of the
ordering process does not depend on the value of  viscosity.
Later, however, the system continues to order only at sufficiently
low viscosities ($\nu \stackrel{<}{\sim} 0.1$). Otherwise, local defects
dominate
  \cite{yeo} and, at times of order  $  t \sim
3000$, the system results frozen in tangled configurations. An
example of evolution at low viscosity ($ \nu = 8.33 \times 10^{-3}
$) is shown in Fig.1. Different kinds of defects can be observed.
The annihilation of a couple of dislocations is put in evidence.
We have measured the relative velocity of the dislocations and we
found it constant. At the last time of the figure the system  can
be observed to be  ordered on larger scales.

A quantitative description of the ordering process comes from the
analysis of  the structure factor $C(\vect{k}, t)$. After spherical
average we plotted $C(k,t)$ at different times. In the early
regime, with $t  \stackrel{<}{\sim} 3000$, $C(k,t)$ develops a maximum at a
momentum $k_M$ which decreases with time until the equilibrium
value $k_0$ is reached. Then the peak of $C(k,t)$, remaining at
$k_0$, continues to grow while the width decreases indicating an
increase of  order in the system. A characteristic length $L$ for
this process can be extracted from the structure factor in the
usual way by measuring the full width $ \delta k$ of
$C$ at half maximum and defining $L(t) = 2 \pi /  \delta k $. In
the inset of Fig.2 we plotted $L(t)$ on log-log scale, calculated
by averaging over 5 different runs. For more than one decade a
behavior consistent with the  power-law  $L \sim t^{z} $ can be
observed with $z = 0.30 \pm 0.02$. The growth becomes slower at
later times and we find $ L \sim \ln t$ as the straight line in
the main frame of Fig.2 suggests. We checked that the slowing down
cannot be attributed to finite size effects (at the latest
times considered, $L(t)$ is less than 1/10 of the size of the
system).

\begin{figure}[t]
\begin{center}
Fig.~3\hspace{0.5\textwidth}Fig.~4
\end{center}
\vskip -0.5cm
\caption{Configuration of a $128^3$ system at $t=1000000$.
The following parameters were used: $a = - b = - 0.026$,
$\kappa = - 0.005$,
$c = 0.0025$, $\Gamma = 0.25$,
and $\nu = 0.1$.}  
\caption{The middle horizontal sections of a $512^3$
systems at times $t=600000$ and $t=700000$. Parameters of this
simulations are the same of Fig.~3.
Observe how the vertical stripes in the central
bottom part of the figures are distorted by the pressure of
horizontal domains.} 
\end{figure}

The behavior of $L$ can be related  to the role played in late
time dynamics by extended defects (grain boundaries) between
domains of differently oriented lamellae. A decrease in the
growing rate of $L$ was observed at times $t\approx 25000$ when
the average value of the angle between the normals of neighboring
domains becomes close to $90^{\circ}$. Isolated  grain boundaries
can be shown to be stable defects in lamellar systems described by
Eq.(\ref{eqn1}) \cite{BV02}.
A quantitative description of
ordering in lamellar systems can be obtained by considering the
evolution equation for a single defect. The position
$\textit{x}_{gb}$ of the grain boundary evolves according to
$\dot{\textit{x}}_{gb} = A \mathcal{C}^2(t) - B \cos(2 k_0
\textit{x}_{gb} + \phi)$ where $ \mathcal{C}$ is the curvature of
lamellae parallel to the boundary and $ A,B,\phi$ are
model-dependent parameters \cite{BV02}.  If one assumes that
dynamical scaling is verified, a typical length can be defined as
the inverse of curvature, and the previous equation implies a
growth like $t^{1/3}$ until a critical value for curvature is
reached. This analysis, elaborated for one single defect, suggests
that in the ordering of a large system, in absence of other
processes and without hydrodynamics, defects will become pinned at
a certain time with the system frozen in a configuration with many
grain boundaries. In our case the velocity field helps the system
in continuing the ordering process also at very late times but
slower.  We measured $L \sim \ln t$ which is the  growth  expected
for systems with metastable states. An extension of the analytical
arguments of \cite{BV02,HV04} in presence of hydrodynamics is not
available. However, it is interesting to observe that our
simulations suggest an intermediate regime with a power law
consistent with that of \cite{BV02} followed by a logarithmic
asymptotic growth, both results explainable as due to grain
boundary dynamics.

We also considered three dimensional systems. 
Simulations ranging from $64^3$ to $512^3$ lattices were performed: The
largest grid asked for about 20 days of sustained computations, using
a 8-CPU vector machine NEC SX6 and more than 32 GB of memory.
While complete order was reached on  $64^3$ lattices,
larger systems confirm the role of grain boundaries 
in lamellar ordering.
An example is given in Fig.3 where a single grain boundary is shown: After 
$t \sim 10^{7}$ no clear 2D ordering was observed.
On the $512^3$ lattice we studied the behavior of $L(t)$,
defined as in  $D=2$. After the formation of lamellae,
the intermediate
and late time  regimes were  found shorter and it was not possible
to deduce  quantitative behaviors. However, the role of grain
boundaries can be appreciated  looking for example at the
configurations of Fig.4. Here  the central horizontal sections of
the system at $t=600000$ and $t=700000$ are shown. Many grain
boundaries with  three  main orientations can be seen. After $t
\sim 600000$, in correspondence of a sudden decrease observed in
the growing rate of $L(t)$, the further evolution of the system
proceeds through the distortion of the grain boundaries existing
at that time. This  can be seen for example in the central bottom
part of the configuration at $t=700000$. Different domains, trying to
increase their size, push themselves reciprocally, helped by the
velocity field. This gives an unexpected late-time increase of
curvature, also observed in $D=2$. The role of hydrodynamics is
important in this process: The system can reach a higher degree of order
only by developing a strong velocity field which bends the existing flat
grain boundaries. This will eventually bring later to a configuration with
less defects. This picture is confirmed by the
increase of kinetic energy observed at $t\sim 600000$.

\section{Dynamical scaling} 
In order to analyze dynamical
scaling, together with $L(t)$, we also considered the behavior of
the peak $C_M \equiv C(k_M,t)$ of the structure factor. In
two-dimensional simulations, from Fig.2, one sees that $C_M$ grows
similarly to $L(t)$ but with a different rate. This suggests to
consider scaling behavior in the general form $C(k,t) \sim
L^{\alpha} f[(k-k_M)L]$ \cite{CRZ94}. This differs from the usual
expression considered in lamellar ordering ($C \sim t^{z}
f[(k-k_M)t^{z}]$) for the introduction of $\alpha$ and
 the use of $L$ even if not given by a power law.
As illustrated in the inset of Fig.5, we evaluated $\alpha= 1.25$.
Then, by using this value, we found, as shown in Fig.5, that
 rescaled structure factors overlap  {\it at all times} after $t \sim 3000$ 
on a single
 curve, confirming
our scaling assumption and  suggesting an analogy with kinetic
behavior in systems with quenched disorder. In Random Field Ising
Models (RFIM), for example,   diffusive growth changes over to
logarithmic behavior  but dynamical scaling holds at all times
with the same scaling function as in  standard Ising model
\cite{RC}.
Here a similar
scenario occurs but without quenched disorder.
This is interesting also in relation with the existence of  an
equilibrium glass transition in systems with lamellar order
\cite{SW00}.

\begin{figure}[t]
\begin{center}
Fig.~5\hspace{0.5\textwidth}Fig.~6
\end{center}
\vskip -0.5cm
\caption{Rescaled structure factors at different times. Symbols
$\triangle, \circ, \star, \bullet, \ast, \diamond$ refer to times
$4600$, $14100$, $40800$, $61100$, $78100$, $123100$,
respectively. In the inset $C_M$ is plotted  on log-log scale as a
function of $L$; the slope of the line is 1.25.}
\caption{Evolution of $L_x$ and $L_y$
for $\dot{\gamma} = 10^{-4} $. The straight line has slope 0.6. At
the latest times of the figure the system is almost completely
ordered, as shown in the $200 \times 200$ portion  of the inset.} 
\end{figure}

Concerning the exponent $\alpha$, its value is related to the
compactness of domains  and is $\alpha = D$   in asymptotic growth
of binary  mixtures, $\alpha = 0 $  in early time regime when
interfaces are forming \cite{CRZ94}. Our results with $\alpha
> 1 $ show that the argument for $\alpha
=1$ based on the existence of one scaling dimension for $2D$
lamellar systems \cite{EVG92} does not hold in presence of
hydrodynamics.
We found  dynamical scaling also in 3D simulations but for a
shorter time interval.

\section{Shear effects}
We also studied lamellar  ordering in presence of  
shear flow.
Shear was applied with bounce-back conditions on  rigid walls
moving with velocity $\pm v_{wall}$ at  the top and  bottom of the
simulation volume \cite{Xu03}. Ordering is favored since
interfaces want  to orientate  with the flow \cite{On97}. Breaking
and recombination of domains, induced by shear,  makes the
elimination of topological defects faster  \cite{prl2}. As a
result, logarithmic growth disappears and ordering occurs  also
asymptotically with power-law behavior  even in cases, at high
viscosity, where without shear freezing would have been observed.

An example of this behavior ($\nu = 7.5$) is shown in Fig.6.
Growth was measured by the quantities $L_{\alpha} = \int d\vect{k}
C(\vect{k},t) / \int d\vect{k} |k_{\alpha}| C(\vect{k},t)$ ($\alpha
=x,y$).
 $L_x$ grows with exponent
$z_x = 0.6$ while $L_y$ relaxes to the equilibrium value.
 Similar results were found for other values of $\nu $;
 a detailed exposition of our results with shear will appear
 elsewhere.

\section{Conclusions}
To conclude, we have studied lamellar ordering in fluid mixtures
with competitive interactions. At low viscosities, when
hydrodynamic modes are effective, dynamical scaling holds in the
form $C(\vect{k} , t) \sim L^{\alpha} f[(k-k_M)L]$ where $L$ is a
characteristic length with complex behavior and logarithmic growth
at late times. This scaling, similar to that  found in systems
with quenched disorder, is shown  here for the first time in 2D
systems with lamellar order. 3D simulations also show the relevance
of grain boundaries  defects for this dynamics.
Shear flow removes extended defects giving power-law behavior at
all times.
\acknowledgments
We warmly thank  F. Corberi, E. Orlandini, M. Zannetti and J.
Yeomans for helpful discussions. We acknowledge
 support by  MIUR (PRIN-2004).


\begin{thebibliography}{99}
\bibitem{det}
Details about the method will appear elsewhere.
\end{thebibliography}
\end{document}